\documentclass[a4paper,amsmath]{jpconf}
\usepackage{graphicx}

\usepackage{slashed}

\begin{document}
\title{Charge Symmetry Breaking  in the Nucleon and Parity Violating Elastic Electron-Proton Scattering}

\author{Gerald A. Miller}

\address{Department of Physics, Univ. of Washington, Seattle, WA 98195-1560 , USA}

\ead{miller@uw.edu}

\begin{abstract}
The basic facts of  charge symmetry breaking (CSB) phenomena are reviewed. The  relevance of CSB to parity-violating electron-proton scattering  experiments that seek to extract strange elastic 
form  factors is discussed. 
Experimentalists have stated and written that the current uncertainty in our knowledge of CSB limits the ability to push further on the strange form factors. I discuss recent calculations using relativistic chiral perturbation theory and realistic values of strong coupling constants which show that the uncertainties due to lack of knowledge of CSB are at least ten times smaller than present experimental uncertainties. 
 Estimates of CSB effects  are made   for the JLab $Q_{\rm weak}$ and Mainz P2 experiments. 
\end{abstract}

\section{Introduction}
This document  is based on review articles~\cite{Henley:1979ig,Miller:1990iz,Miller:1994zh,Miller:2006tv}, which describe  diverse aspects of charge symmetry and its breaking, and a recent
article with student Michael Wagman~\cite{Wagman:2014nfa}.  Unpublished results related to the $Q_{\rm Weak}$~\cite{Androic:2013rhu} and P2 experiments are also presented.

Charge symmetry CS  is invariance under an isospin  rotation of $\pi$ about the $y$-axis in isospin space. Thus a $u$ 
  quark is rotated into a $d$ quark. CS is broken slightly by the light-quark mass difference and by electromagnetic effects.
  Isospin invariance, or $[H,T_i]=0$ is invariance under all rotations in isospin space. This invariance  is also called charge independence, CI. 
  
  Charge symmetry does not imply isospin invariance.  
  For example, the mass difference between charged and neutral pions breaks isospin invariance but not charge symmetry. Another example is the measured difference between the $^1S_0$ $np$ and $nn$ scattering
  lengths.
  
  In general the size of CSB effects  is much smaller than the breaking of isospin invariance, CIB. The 
  scale of CSB is typified by the ratio of the neutron-proton mass difference to the proton mass which is about one part in 1000. This is much smaller than the pion mass difference effect which is one part in 27. The CIB of nucleon-nucleon scattering lengths was discovered well before 1965,  but unambiguous evidence for   CSB in nucleon-nucleon scattering  did not appear  until about  1979.
  Thus the expectation is that CSB is a small effect, uncovered only with special effort. The small relative  size of CSB effects compared with those of CIB is a consistent with the power counting of
  chiral perturbation theory~\cite{vanKolck:1996rm}.

The IUCF measurement of the nonzero cross section for the reaction $dd\to \alpha\pi^0$ is a noteworthy example of CSB~\cite{Stephenson:2003dv}.
The $d$ and $\alpha$ are even under the CS transformation, while the $\pi^0$ is  odd under that transformation. Thus the observed cross section is a measurement of a {\it square}  of a symmetry-violating matrix element. The size of the measured cross section ($\sim 14$ pb) is in qualitative agreement with 
predictions based on a hybrid form of chiral perturbation theory~\cite{Gardestig:2004hs,Nogga:2006cp}.

\section{Parity Violating Electron Scattering (PVES), Strangeness Electromagnetic Nucleon Form Factors and Charge Symmetry Breaking}
 Understanding parity-violating electron scattering requires knowledge of weak neutral form factors. These are sensitive to nucleon strangeness content and
 the value of the weak mixing angle. The latest review is that of Armstrong \& McKeown~\cite{Armstrong:2012bi} who conclude that a  convincing signal for nucleon strangeness has not been seen.  A close look at the data shows that the  uncertainty around 0 in the strangeness magnetic form factor, $G_M^s$ is of the order of about 0.3 nucleon magnetons, which is not that small. The uncertainty around 0  on $G_E$ is not so small either.

The relevance of charge symmetry to PVES is that charge symmetry is used to express the  form factor for $Z$-boson exchange
in terms of measured electromagnetic form factors.
One finds the result
\begin{equation}
 G_{E,M}^{Z,p} = (1 - 4 \sin^2 \theta_W) G_{E,M}^{\gamma,p}
		- G_{E,M}^{\gamma,n} - G_{E,M}^{s}, \label{eq:EMZ}
\end{equation}
where $G_{E,M}$ are Sachs form factors by assuming that $u$ in proton corresponds to $d$ in the neutron, $d$ in proton corresponds to  $u$ in the neutron.

However, if one includes the possibility that CS is violated one finds
\begin{equation}
 G_{E,M}^{Z,p} = (1 - 4 \sin^2 \theta_W) G_{E,M}^{\gamma,p}
		- G_{E,M}^{\gamma,n} - G_{E,M}^{s} -G^{\rm CSB}(Q^2), \label{eq:EMZ}
\end{equation}
with the CSB term $G^{\rm CSB}(Q^2)$ computed from the expression
\newcommand{\mbraket}[3]{\left< #1 \vphantom{#2#3} \right|
 #2 \left| #3 \vphantom{#1#2} \right>} 
 \newcommand{\eq}[1]{Eq.~(\ref{#1})}
\begin{eqnarray}
	\label{defFvs}
  \bar{u}(p^\prime)\left[ \gamma^\mu F_1^{\rm CSB}(Q^2) + \frac{i\sigma^{\mu\nu}q_\nu}{2m_N}F_2^{\rm CSB}(Q^2) \right]u(p)&=&  \mbraket{p(p^\prime)}{\frac{1}{3}\bar{u}\gamma^\mu u - \frac{2}{3}\bar{d}\gamma^\mu d}{p(p)}\\\nonumber\\\nonumber
	&&\hspace{10pt} + \mbraket{n(p^\prime)}{\frac{2}{3}\bar{u}\gamma^\mu u - \frac{1}{3}\bar{d}\gamma^\mu d}{n(p)}.
\end{eqnarray}
The charge symmetry transformation rotates the first term on the right-hand-side of \eq{defFvs}  into the negative of the second term, so that if CS holds
$F_{1,2}^{\rm CSB}=0$.
But CS is known to be violated, so one needs to know about the relative size of the terms $G_{E,M}^{\rm CSB}$ and $G_{E.M}^s$.

The   worry about a possibly large theoretical uncertainty  in $G_{E,M}^{\rm CSB}$  has caused  experimentalists to stop their efforts to discover strangeness in nucleons through elastic electromagnetic form factors. For example
Ref.~\cite{Acha:2006my} states ``Theoretical uncertainties especially regarding the assumption of charge symmetry [24], preclude significant improvement to the measurements reported here."
(Ref. [24] of  \cite{Acha:2006my} is our Ref.~\cite{Kubis:2006cy}.) Similar remarks are made in ~\cite{Paschke:2011zz}. However, Ref.~\cite{Wang:1900ta} states that  the charge symmetry effect ``estimated in the calculation of Kubis and Lewis [53] (our Ref.~\cite{Kubis:2006cy}) is an exception" to the general experience that charge symmetry  breaking effects being very small and that 
``implications of this work [53] for other examples of charge symmetry violation have not yet been worked out."  The statements of Ref.~\cite{Wang:1900ta} originate from the strong vector-meson nucleon coupling constants that Kubis \& Lewis employ in their resonance saturation procedure. These coupling constants are a focus of the present work. Also note that   Ref.~\cite{GonzalezJimenez:2011fq} simply states ``isospin violations ... are expected to be very small." So there seems to be a divergence of opinion regarding the importance  of the charge symmetry breaking effects. Given the  large interest in the strangeness content of the nucleon, it is of considerable relevance  to re-examine the charge symmetry breaking effects, and we    do that here.

\section{Some history of CSB in PVES}
I studied this problem long ago~\cite{Miller:1997ya}  and found, using a set of SU(6) non-relativistic quark models, that the effects of the charge symmetry breaking  are less than about 1\%  for experimentally relevant values of the momentum transfer.
The 1\% refers to the ratio of $G_{E.M}^{\rm CSB}$ to the measured $G_{E,M}$. The relevant standard now is the size of $G_{E,M}^{\rm CSB}$ relative to the experimental uncertainties.  The effects of the up-down quark mass difference in the kinetic energy and one-gluon exchange potential energy along with electromagnetic effects give tiny contributions, especially at low values of $Q^2$.

However, I had left out the effects of the neutron-proton mass difference ($M_n-M_p=\Delta M$) that appear in evaluating the contributions of the pion cloud to electromagnetic form factors. The proton occasionally fluctuates into a neutron and a $\pi^+$ and a neutron occasionally fluctuates into a proton and a $\pi^-$.   These are not exactly the same because of  the neutron-proton mass difference. Kubis \& Lewis~\cite{Kubis:2006cy} found the effect using  heavy baryon chiral perturbation theory (HB$\chi$PT).
It turns out that the presence of a chiral logarithmic divergence, $\sim \log (\Lambda/m_\pi)$,  with $\Lambda$ large, gives an enhancement that outweighs the very small value of 
$\Delta M$.

\begin{figure}
  \includegraphics[width=.8\textwidth]{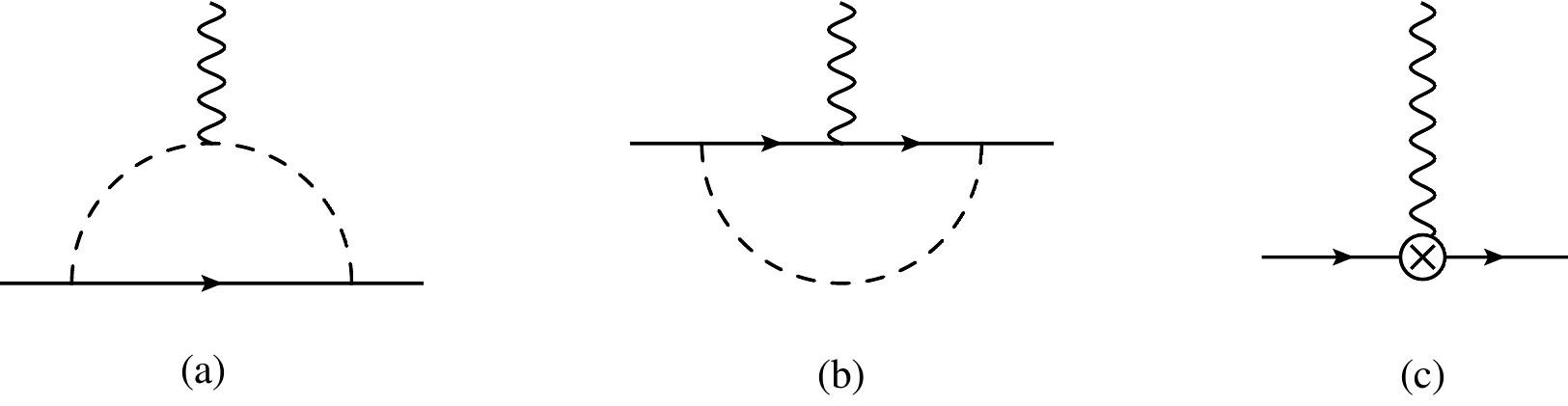}
	\caption{The leading CSB contributions to the proton's neutral weak form factors in chiral perturbation theory. CSB effects arise in the pion loop diagrams (a) and (b) from the proton-neutron mass difference. The crossed circle in diagram (c) represents a CSB nucleon-photon interaction  
	arising from short distance interactions that arises in chiral perturbation theory. Wave function renormalization also gives a CSB contribution not shown.} 
	\label{diagrams}
\end{figure}

But this use of HB$\chi$PT  has a problem. One-pion-exchange contributions computed  shown diagrammatically in Fig.~\ref{diagrams}, do not lead to an unambiguous   prediction for the CSB contribution to the neutral weak magnetic form factor   because a counterterm unconstrained by symmetry or experiment contributes at leading-order (LO) in chiral power counting.

A prediction based upon $\chi$PT requires a model estimate of this unconstrained counterterm. Kubis \& Lewis (KL)~\cite{Kubis:2006cy} used a resonance saturation technique in which the CSB is driven by $\rho-\omega$ mixing,  to provide such a model estimate. 
The physical states of the $\rho$ and $\omega$ mesons are actually not pure states of isospin because of the $u-d$ mass difference and because of electromagnetic effects. The net result of KL is obtained by 
combining the estimate of the effects of $\rho-\omega$ mixing  with calculations in HB$\chi$PT and infrared regularized baryon chiral perturbation theory. KL predicted a CSB magnetic moment contribution of $0.025\pm 0.015$ including resonance parameter uncertainty~\cite{Kubis:2006cy}. This effect is an order of magnitude smaller than current experimental uncertainties in nucleon strangeness measurements, but it is larger than predictions based on non-relativistic quarks models or naive dimensional analysis.

The effects of $\rho-\omega $ mixing graphs kill the infinity and provide a finite contribution, which is larger~\cite{Kubis:2006cy}   than the pion loop diagram, and is scale-dependent. Thus an ambiguity remains. The KL results for $G_E$ are similar to my early results. Their results for 
$G_M^{\rm CSB}\approx 0.02\pm 0.015$ are much larger but still very small, about an order of magnitude smaller than the experimental uncertainty.  The next-to-leading order term gives a 100 \% correction, so the calculation is not converged.  The large spread in the computed values of $G_M^{\rm CSB}$ arises from  the uncertainty in the strong tensor  coupling of the $\omega$ to the nucleon, which  is generally expected to be small. This surprising result arises because
KL take strong coupling constants form dispersion analyses of electromagnetic form factors and therefore use strong $\omega-N$ coupling  constants are much, much larger than typically used in nucleon-nucleon scattering.  Therefore there were two new aspects that we considered in our paper~\cite{Wagman:2014nfa}. The first was to confirm the appearance of a chiral log term, and the second was to study and constrain the values
of the strong coupling constants.
\newcommand{\fs}[1]{\slashed{#1}} 

\section{Formalism of   Wagman \& Miller~\cite{Wagman:2014nfa}}

We use the chiral Lagrangian 
 \begin{eqnarray}
  \mathcal{L}_{N\pi\gamma} &=& \bar{N}\left[ i\fs{\partial} -Q\fs{A} - \left( m_N - \frac{\Delta m_N}{2}\tau_3 \right) - \frac{g_A}{2f_\pi}\partial_\mu \pi^a \gamma^\mu \gamma_5 \tau^a\right.\label{LNpigamma}
\left.+ \frac{e\sigma^{\mu\nu}}{4m_N}F_{\mu\nu}(\kappa^{\fs{v}} + \kappa^{\fs{s}}\tau^3)  \right]N,
	\end{eqnarray}
along with pion kinetic energy terms, appropriate for relativistic chiral perturbation theory (RB$\chi$PT). In this $N $ is an isospinor for the nucleon fields, $Q$ is the nucleon charge matrix, $A^\mu$ and $F^{\mu\nu}$  are the usual photon field and field strength tensor, $m_N = 938.9$ MeV is the average nucleon mass, $g_A = 1.2701(25)$ is the axial charge of the nucleon, $f_\pi = 92.21(14)$ MeV is the pion decay constant, $\pi^a$ is an isovector of pion fields, the $\tau^a$ are Pauli matrices acting in isospin space, and $\Delta m_N = m_n - m_p = 1.2933322(4)$ MeV is the nucleon mass splitting.  The final term is a nucleon-photon contact interaction allowed by symmetries and power counting (quark model stuff and vector meson contributions).

We use relativistic chiral PT and compute the relevant graphs using usual Feynman rules without the need to expand in inverse powers of the nucleon mass. More equations can be found in~\cite{Wagman:2014nfa}.  We found that our relativistic $F_2^{\rm CSB}$ is a convergent integral that has a scale-independent chiral log term  of the form $\log m_N/m_\pi$ with the same coefficient as KL.  This contrast with the chiral log of KL which is cut off at the rho meson mass and has a scale-dependent counter term.

We also include the effects of $\rho-\omega$ mixing. But the coupling constants are constrained because these terms appear in nucleon-nucleon scattering as an NN potential of medium range. The effects of $\rho-\omega$ have been  reviewed~\cite{Henley:1979ig,Miller:1990iz,Miller:1994zh,Miller:2006tv}.  This term gives contributions to the $nn$ vs $pp$ $^1S_0$ scattering length, and cause class IV forces~\cite{Henley:1979ig}
in $np$ scattering. This force causes a polarization difference between the outgoing neutron and proton. This effect contributes significantly to
the binding energy difference between $^3H$ and $^3He$ and in other mirror nuclei.  The $\rho-\omega$ mixing effects do not dominate all of the mentioned quantities, but contributes to all. However,  all known strong contributions to CSB have the same sign because they are all driven by the same up-down mass quark difference. The point is that the relevant coupling constants are constrained by existing knowledge.

\section{Results  of  Wagman \& Miller~\cite{Wagman:2014nfa}}
Please see the cited reference to find a full description of our results. Here there is only space available to summarize. The value of $G_E^{\rm CSB}(Q^2)$ vanishes at $Q^2=0$ and rises monotonically  to a maximum of about $0.001 \pm 0.0005$  at $Q^2=0.3$ GeV$^2$.
This is small indeed.  Our best estimate of $G_M$ is that it is approximately constant at $0.01 \pm 0.01$ for $Q^2$ between 0 and 0.3 GeV$^2$.

The predictions for the CSB magnetic moment and electric and magnetic radii are shown in Table~\ref{moments}. Strictly speaking only the $Q^2 = 0$ resonance contribution should be counted as LO, but the higher-order contributions found by including the full $Q^2$ dependence of $F_1^{CSB}(Q^2)_{\rho-\omega}$ and $F_2^{CSB}(Q^2)_{\rho-\omega}$ are numerically significant and we include them in our results. The loop contributions are also shown in Table~\ref{moments}, both with and without phenomenological form factor effects.

\begin{table}
	\centering
	\caption{ Results for the CSB magnetic moment $G_M^{CSB}(0)$ and electric and magnetic radii $\rho^{CSB} = -\frac{dG^{CSB}}{dQ^2}(0)$. The two lines show the loop contributions only, without and with phenomenological $\pi\pi\gamma$ and $\pi NN$ vertex form factors FF. The third and fourth lines show our full LO predictions for the form factor moments with the unconstrained counterterm $\kappa_{CT}^{CSB}$ estimated with resonance saturation. The uncertainties shown are from uncertainties in the resonance parameters $\kappa_\omega$ and $\Theta_{\rho\omega}$. See ~\cite{Wagman:2014nfa} for further explanations and results. }
	\begin{tabular}{|l|| c| c | c|}\hline
	 & $G_M^{CSB}(0)$ & $\rho_M^{CSB}$ (fm$^2$) & $\rho_E^{CSB}$ (fm$^2$)\\\hline
	Loop& 0.014 & 0.012 & 0.0004 \\\hline
	Loop + FF & 0.006 & 0.0017 & 0.0009 \\\hline
	Loop + Resonance, $g_\omega$ = 10 & 0.0.019 $\pm$ 0.003& 0.001 $\pm$ 0.0004 & -0.003 $\pm$ 0.0001 \\\hline
	Loop + Resonance + FF, $g_\omega=10$ &0.012 $\pm$ 0.003 & 0.0006 $\pm$ 0.0004 &-0.0007  $\pm$0.0001\\\hline
\end{tabular}

	\label{moments}
\end{table}
\section{Effect on  the $Q_{\rm weak}$~\cite{Androic:2013rhu} and P2 experiments}
 The formula that relates the CSB form factors to the observed experimental asymmetry is
\begin{eqnarray}
\delta A^{\rm PV}_{\rm CSB}=-A_0\frac{\epsilon G_{ E}^{\gamma  p}G^{\rm CSB}_E+\tau G_{ M}^{\gamma  p}G^{\rm CSB}_M}
{\epsilon (G_{ E}^{\gamma  p})^2+\tau (G_{ M}^{\gamma  p})^2},
\end{eqnarray}
where $\epsilon\approx 1$ and $\tau\approx 0.07$ for the $Q_{\rm weak}$ experiment. This means that electric effects are dominant. Since $G_E^{\rm CSB}$ is very small we find
$|\delta A^{\rm PV}_{\rm CSB}|<1$ ppb. This is negligible compared with the observed 279 ppb. The Mainz P2 experiment is planned to run at even smaller values of $Q^2$ and our estimate for this experiment is 
an even smaller value of $|\delta A^{\rm PV}_{\rm CSB}|<0.01$ ppb.  The absolute value symbols  appear because $G_E^{\rm CSB}$ could be negative.

\section{Summary   of Miller \& Wagman~\cite{Wagman:2014nfa}}
Our principal result is that charge symmetry breaking effects are too small to influence the extraction of nucleon strangeness measurements from parity-violating electron-proton scattering experiments. Including both uncertainty in resonance parameters and higher-order term uncertainty quantified by the magnitude of form factor contributions, our LO predictions are $G_M^{CSB}(0) = 0.021 \pm 0.01 \pm 0.008$ and $|G_E^{CSB}| < 0.005$ for $Q^2 < 0.3$ GeV$^2$. Comparing these results with current experimental bounds on strangeness form factors $G_M^s = 0.33 \pm 0.4$, $G_E^s = 0.006 \pm 0.02$ at $Q^2 = 0.1$ GeV$^2$~\cite{Armstrong:2012bi}, we see that our CSB predictions are an order of magnitude smaller than current experimental error bars. 

The predictions of HB$\chi$PT with resonance saturation made by KL are  $G_M^{CSB}(0) = 0.025 \pm 0.02$ and $|G_E^{CSB}| < 0.01$ for $Q^2 < 0.03$ GeV$^2$ including resonance parameter uncertainty~\cite{Kubis:2006cy}. The much larger resonance parameter uncertainty in these results arises from using a large $\omega$-nucleon coupling constant $g_\omega \sim 42$ taken from dispersion analysis. Experimental measurements of the $^3$He-$^3$H binding energy difference constrain $g_\omega$ to be less than about $19 \pm 5$ when $\rho-\omega$ mixing is treated as a resonance contribution to HB$\chi$PT contact operators. Taking $g_\omega = 19$, the prediction of HB$\chi$PT with resonance saturation becomes $G_M^{CSB}(0) = 0.031 \pm 0.01$ at NLO and $|G_E^{CSB}|<0.005$ at LO for $Q^2 < 0.03$ GeV$^2$. This is once again an order of magnitude smaller than current experimental uncertainties on nucleon strangeness.

Our results demonstrate good agreement between LO loop contributions in RB$\chi$PT and HB$\chi$PT. The RB$\chi$PT loop contribution of 0.014 to $G_M^{CSB}(0)$ agrees with the LO HB$\chi$PT loop contribution to better than 95\%. The RB$\chi$PT loop contribution to $\rho_M^{CSB}$ is smaller than the LO HB$\chi$PT loop contribution but larger than the loop contribution at NLO. The two frameworks therefore manifestly agree on $\rho_M^{CSB}$ up to higher-order corrections. The RB$\chi$PT loop contribution to $\rho_E^{CSB}$ is also smaller than the LO HB$\chi$PT loop contribution, but $\rho_E^{CSB}$ is numerically dominated by the resonance contribution in both frameworks and so we expect that differences can again be considered higher-order. 

RB$\chi$PT and HB$\chi$PT must give predictions for physical observables that agree up to higher-order errors once loop and counterterm contributions are included. It is encouraging to see that this agreement is achieved when using resonance saturation estimates for the counterterm contributions. A model independent chiral prediction for the CSB form factors still requires direct constraints on the counterterm contribution from experiment or QCD, but our investigations have found no reason to doubt the consistency of CSB form factor predictions using chiral loops and resonance saturation counterterms.  Therefore, we may conclude that the  theoretical uncertainties are  under control.

\section*{Acknowledgments}
This material is based upon work supported by the U.S. Department of Energy Office of Science, Office of Basic Energy Sciences program under Award Number DE-FG02-97ER-41014.

 \section*{References}
\bibliography{csb}

\providecommand{\newblock}{}
\begin{thebibliography}{10}
\expandafter\ifx\csname url\endcsname\relax
  \def\url#1{{\tt #1}}\fi
\expandafter\ifx\csname urlprefix\endcsname\relax\def\urlprefix{URL }\fi
\providecommand{\eprint}[2][]{\url{#2}}

\bibitem{Henley:1979ig}
Henley E and Miller G 1979 {\em Mesons in Nuclei\/} vol~1 (Amsterdam:
  North-Holland/Asterdam) p 405

\bibitem{Miller:1990iz}
Miller G, Nefkens B and Slaus I 1990 {\em Phys.Rept.\/} {\bf 194} 1--116

\bibitem{Miller:1994zh}
Miller G~A and Oers W~T~H~V 1995 {\em Symmetries and Fundamental Interactions
  in Nuclei\/} (Singapore: World Scientific) p 127

\bibitem{Miller:2006tv}
Miller G~A, Opper A~K and Stephenson E~J 2006 {\em Ann.Rev.Nucl.Part.Sci.\/}
  {\bf 56} 253--292

\bibitem{Wagman:2014nfa}
Wagman M and Miller G~A 2014 {\em Phys.Rev.\/} {\bf C89} 065206

\bibitem{Androic:2013rhu}
Androic D {\em et~al.\/} (Qweak Collaboration) 2013 {\em Phys.Rev.Lett.\/} {\bf
  111} 141803

\bibitem{vanKolck:1996rm}
van Kolck U, Friar J~L and Goldman J~T 1996 {\em Phys.Lett.\/} {\bf B371}
  169--174

\bibitem{Stephenson:2003dv}
Stephenson E, Bacher A, Allgower C, Gardestig A, Lavelle C {\em et~al.\/} 2003
  {\em Phys.Rev.Lett.\/} {\bf 91} 142302

\bibitem{Gardestig:2004hs}
Gardestig A, Horowitz C, Nogga A, Fonseca A, Hanhart C {\em et~al.\/} 2004 {\em
  Phys.Rev.\/} {\bf C69} 044606

\bibitem{Nogga:2006cp}
Nogga A, Fonseca A, Gardestig A, Hanhart C, Horowitz C {\em et~al.\/} 2006 {\em
  Phys.Lett.\/} {\bf B639} 465--470

\bibitem{Armstrong:2012bi}
Armstrong D and McKeown R 2012 {\em Ann.Rev.Nucl.Part.Sci.\/} {\bf 62} 337--359

\bibitem{Acha:2006my}
Acha A {\em et~al.\/} (HAPPEX collaboration) 2007 {\em Phys.Rev.Lett.\/} {\bf
  98} 032301

\bibitem{Kubis:2006cy}
Kubis B and Lewis R 2006 {\em Phys.Rev.\/} {\bf C74} 015204

\bibitem{Paschke:2011zz}
Paschke K, Thomas A, Michaels R and Armstrong D 2011 {\em J.Phys.Conf.Ser.\/}
  {\bf 299} 012003

\bibitem{Wang:1900ta}
Wang P, Leinweber D, Thomas A and Young R 2009 {\em Phys.Rev.\/} {\bf C79}
  065202

\bibitem{GonzalezJimenez:2011fq}
Gonzalez-Jimenez R, Caballero J and Donnelly T 2013 {\em Phys.Rept.\/} {\bf
  524} 1--35

\bibitem{Miller:1997ya}
Miller G~A 1998 {\em Phys.Rev.\/} {\bf C57} 1492--1505

\end{thebibliography}
\end{document}